# Sources of Change for Modern Knowledge Organization Systems

Michael Lauruhn and Paul Groth
{m.lauruhn, p.groth}@elsevier.com
Elsevier Labs
1600 John F. Kennedy Boulevard, Suite 1800,
Philadelphia, PA

## Abstract
Knowledge Organization Systems (e.g. taxonomies and ontologies) continue to contribute benefits in the design of information systems by providing a shared conceptual underpinning for developers, users, and automated systems. However, the standard mechanisms for the management of KOSs changes are inadequate for systems built on top of thousands of data sources or with the involvement of hundreds of individuals. In this work, we review standard sources of change for KOSs (e.g. institutional shifts; standards cycles; cultural and political; distribution, etc) and then proceed to catalog new sources of change for KOSs ranging from massively cooperative development to always-on automated extraction systems. Finally, we reflect on what this means for the design and management of KOSs.

## 1. Introduction

Knowledge Organization Systems (KOSs) such as ontologies, terminologies, data dictionaries, and classification schemes provide the foundation for a variety of applications. These applications range from classification of objects, indexing processes, and traditional information retrieval (IR) systems to semantic web applications, question answering, and rule-based systems. While the core goal of many KOSs is to resolve entities and concepts for the applications they serve, newer functionality includes reasoning and discovery. Traditionally, KOSs have depended on manual processes that were largely akin to an editorial process based entirely on human supervision, recommendation, and decisions. These were typically done by, or with input and influence from, subject matter experts or collection experts. As KOSs creation and maintenance processes evolve, one significant change is the level of human input that can effect change.

We note how sources of change differ as KOSs evolve across various models over time. We begin by looking at issues related to evaluating and correcting bias in early KOSs (Library of Congress Subject Headings and Dewey Decimal Classification) that were designed for library cataloging. These models had a wide breadth in scope and regularly scheduled distribution processes, initially exclusively in print. These earlier KOS rely heavily on full-time editors and dedicated subject matters and the development and dissemination process can be equated to an editorial system the produces editions on a regular schedule. Over time, KOSs with scopes that are more niche became more prevalent. In this paper we will examine a pair that are created for indexing biomedical literature and databases. The editorial processes are still based on human decision making, but begin to come from a more distributed group of experts who are both users of and contributors to the model. Likewise, the sources of change also begin to become more application focused. As the contributor model becomes more distributed, the release of updates and change evolve into an ongoing process with more frequent versions that start to resemble a software release cycle.

In this work, our aim is to provide a catalog of sources of change that designers and users of KOSs need to be aware of. Figure 1 depicts the nine source of change we call out in this paper and how they map into the major constitutes that impact the maintenance and creation of KOSs.

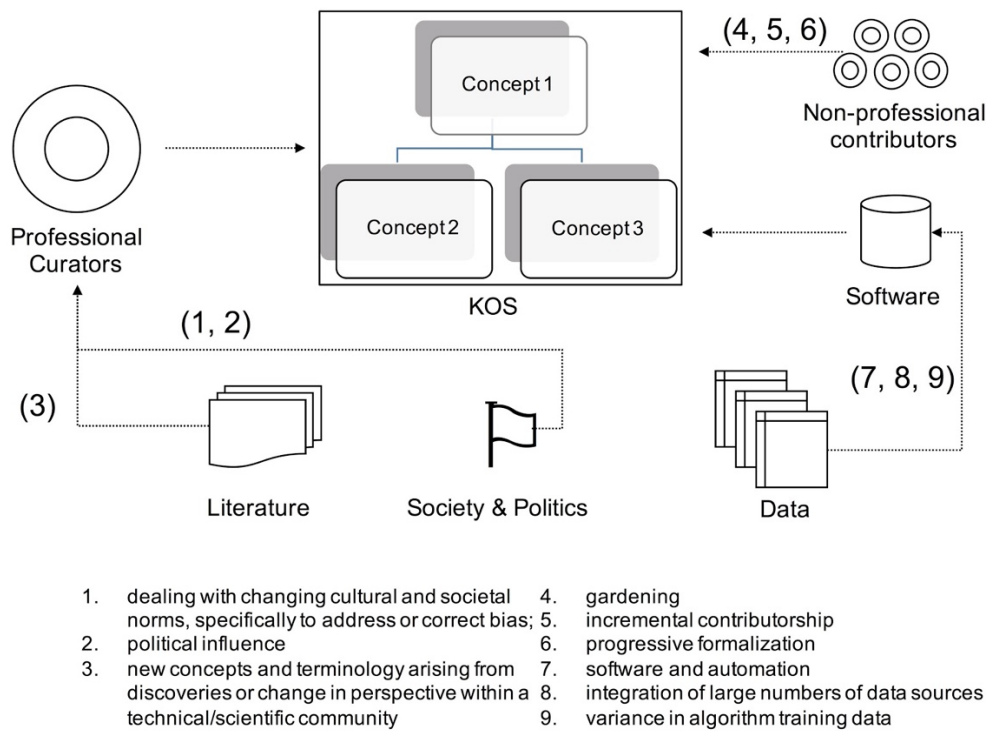

1. dealing with changing cultural and societal norms, specifically to address or correct bias;
2. political influence
3. new concepts and terminology arising from discoveries or change in perspective within a technical/scientific community
4. gardening
5. incremental contributorship
6. progressive formalization
7. software and automation
8. integration of large numbers of data sources
9. variance in algorithm training data

Figure 1: Sources of changed mapped to the major actors and entities within KOS creation and maintenance processes.

To substantiate this list, we begin by a review of the existing literature. We then proceed to catalog new sources of change with appropriate exemplars. This is followed by a discussion about the implications all these sources of change have on KOS design and management.

## 2. Sources of KOS Change: The Existing View
### 2.1 Change as an editorial process

One of the major drivers behind change in KOSs, particularly ones that are largely-distributed across disciplines, is a desire to remain up to date in terms of cultural sensitivities and perceived bias. In his 1971 book, *Prejudices and Antipathies: A Tract on the LC Subject Heads Concerning People*, Sanford Berman outlines many of the subject areas in which the Library of Congress Subject Headings (LCSH) are insensitive and out of touch with the society that libraries operated in (Berman 1971). These include areas such as race and ethnic groups, gender roles in society, politics, sexuality, and others. In the 1971 introduction, Berman cites previous literature which identifies that earlier justifications for the bias that was prevalent in the LCSH. The author then encourages others to join him in seeking to "remedy long-standing mistakes and to gain for the profession a genuine, earned respect among people who read and think."

In his subsequent 1993 Preface to *Prejudices and Antipathies*, Berman reflected on some of the changes that LCSH had undergone since the book was originally published: "But these changes [that had occurred], however welcome, are no cause for gloating. It took 13 years for LC to scrap JEWISH

QUESTION and 18 to eliminate YELLOW PERIL, hardly examples of swift response and profound sensitivity" (Berman 1993).

Eleven years later, in 2004, (Knowlton 2004) published a study that compiled Berman's suggestions and tracked the changes that had occurred. It found that 39% of the subjects were changed in a manner that nearly matches Berman's suggestions, while 36% were unchanged, and 24% changed in a manner that partially captures the suggestions. Knowlton acknowledges that compilers of LCSH had addressed bias in a serious manner in the years following Berman's original work. He notes that many of the suggestions that went unchanged "simply reflect a difference of opinion on the literary merit of subject headings changes" while elsewhere large swaths may reflect continuing bias.

To outside observers, it might seem that making changes in order to remain up to date on cultural and societal norms may make common sense and *not* making changes borders on egregious and insensitive behavior. However, even if an organization wants to make changes, often times those changes may be difficult to execute due to processes, rationale, and decision-making required to reach consensus and redefine structures in the model. These delays can be further extended when the KOS is traditionally released on a regular basis and there is a reluctance to make changes that may have to be changed again or changed back in the future. As an example, (Green 2015) highlights many of the issues dealing with how Indigenous Peoples in the United States are represented in the Dewey Decimal Classification (DDC). Complexities include a many-to-many relationship that exists between ethnic groups and federally recognized tribes in the United States. In addition, while there maybe a concept recognizing an ethnic group, that group is not represented as sovereign nation. In addition, there may be some confusion as some nations are represented as geographic concepts. In acknowledging some of the claims of bias against the DDC, Green notes that "some claims are perhaps based on misunderstanding, but some point to areas where the DDC can be improved." Green also illustrates that the decision-making process that leads to eventual change will include a vetting with the indigenous communities in question. This is an effort so that DDC's classification principles and practices are upheld, while being "true to the voice and perspective of the peoples being represented."

Another source for change and, in some cases, influences *against* making desired changes to a KOS come from stakeholders not directly affiliated with the day-to-day maintenance of a KOS and their constituencies. In the winter of 2014, a student group from Dartmouth College began a movement to remove "illegal aliens" as a term in the LCSH (Qin 2016). The grassroots efforts of the Coalition for Immigration Reform, Equality and DREAMers (CoFIRED) resulting in a petition being presented to the Library of Congress in the summer of 2014. In 2016, the Library of Congress made clear its intentions to remove the term "Illegal Alien" from LCSH and replace it with "noncitizen" and "unauthorized immigration" (Aguilera 2016). This change was met with objections from members of the United States Congress, with members going as far as adding a clause to an Appropriations Bill (https://www.congress.gov/congressional-report/114th-congress/house-report/594/1) that instructs "the Library [of Congress] to maintain certain subject headings that reflect terminology used in title 8, United States Code," and subsequently leaving the terminology in tact. [*Note:* At the time of this writing in July 2016, the Library of Congress plans to stop using the term "illegal alien," as they are also receiving public comment on the matter and the passage of the related Legislative Branch Appropriations Bill is pending.]

As a means to create a larger community of contributors, the Library of Congress has formal channels to allow for a larger group of organizations and institutions to make proposals for changes in LCSH. The Subject Authority Cooperative (SACO) program uses the metaphor of the funnel to illustrate ideas and topics move through a system of deliberation and professional judgement. Funnels are groups of libraries or catalogers that work in subject areas or specific regions and have joined together to contribute subject authority records for inclusion in the Library of Congress Subject Headings.

The examples above show widely-distributed, cross-disciplinary vocabularies that have been traditionally distributed in regular releases of volumes and editions, where it is not desirable to roll-back changes, thus significant changes are likely to be deliberated over time. It is worth noting that each of the previous examples deal with KOS that include people, whether as ethnic groups, races, populations, or various other groups as part of their subject matter. The changes illustrated to the KOSs are based upon desires to correct previous biases that exist in the models.

## 2.2 Application-specific sources of change

Other KOSs have been more niche in their scope and faced different sources of change. Because of their specialized mandate, audience, and governance, they have different change processes. In many cases, the most common change for these KOSs is the addition of new concepts that are discovered within their domain. For the most part, these types additions are not controversial and the processes depart from large interdisciplinary editorial-style deliberation to one where decision-making is left largely to dedicated subject matter experts.

Medical Subject Headings (MeSH) is the United States National Library of Medicine's thesaurus. MeSH has section staff members who are responsible for ongoing revisions to the MeSH vocabulary. The MeSH website identifies three sources for changes: Subject specialists make changes in the areas of their expertise, indexers and others may make suggestions to the subject specialists, and the staff collect new concepts and terminology from scientific literature and emerging research (U.S. National Library of Medicine, "MeSH Fact Sheet" 2015).

In describing their approach to decisions about changes to MeSH, the NLM describes an approach that focuses on the primary use cases for the model, indexing and cataloging scientific literature. There seems to be a stronger emphasis on usefulness toward those tasks than there is for completeness from a strict ontological viewpoint. "There are many factors that must be considered in deciding whether to add a MeSH descriptor. An interest in one species of a given genus, may lead to interest in some other species or even all of that genus. Yet, if there is little published about the other species, there is little purpose or advantage in creating a myriad of new descriptors in a vocabulary designed to describe the subject content of published literature. Before new descriptors are introduced, there is careful consideration of how the concept is currently indexed or cataloged. If the existing descriptors and qualifiers (subheadings) precisely characterize or identify the literature on the subject, there may not be a need for a new descriptor. Both too much change or too little change are to be avoided as MeSH is kept current with changes in biomedical knowledge" (U.S. National Library of Medicine, "MeSH Vocabulary Changes" 2015).

In 2014, the NLM formed the Linked Data Infrastructure Working Group to investigate determine best practices for publishing linked data (Bushman et al 2015). As part of the initiative, publishing MeSH in RDF was selected as a Linked Data pilot. The first beta version of MeSH RDF was based on the 2014 version of the vocabulary. When the next beta release of MeSH in RDF was released in June 2015, it was based on the 2015 version of MeSH. However this marked a significant change, as moving forward NLM was able to make daily updates. MeSH in RDF has since moved out of beta and the daily update process is able to capture "emergency updates to MeSH." As an example, in spring of 2016, descriptors for Zika Virus and Zika Virus Infections were added as Zika Virus epidemic spread in South America. Those changes will be incorporated in an annual version in the fall, when the static graph for the 2016 version of MeSH is generated.

## 2.3 Distributed stakeholders

EMTREE is a proprietary thesaurus created by Elsevier in order to support indexing of EMBASE, the company's database of literature in the biomedical domain. At the beginning of 2016 the thesaurus contained more than 73,000 preferred terms and more than 310,000 synonyms (Elsevier, "Embase Content").

In "Change Management for Distributed Ontologies," (Klein 2004) studied change processes for EMTREE and documented another largely human-directed process. Similar to MeSH, it was noted that most changes were to account for new terms and concepts and that the EMTREE users suggested. Users indexing articles suggest additions throughout the year and accepted additions first go to a development version and later (between 3 and 15 months) appear in a production version.
Also similar to MeSH, decisions to add concepts are based largely upon frequency that they appear in literature and that the two most common reasons for not including new concepts is that the information is either incorrect or does not occur enough. EMTREE also contains mapping to MeSH, so another form a change that must be accounted for changes introduced in MeSH.

Another example from the sciences shows a more niche topic and a more distributed source of inputs from people. The Gene Ontology (GO) project is a bioinformatics project that builds and maintains ontologies for more than 40,000 biological concepts. The primary use of the ontology is to represent concepts used to annotate experiments that feature gene functions as reported in scientific articles and papers. The ontology is in a constant state of change to capture new discoveries (http://geneontology.org/page/about).

Klein describes a process where a small number of full time curators work on the vocabulary and its relations, but relies on GO users to make suggestions for new terms or edits. A change request system allows the those users to track their submissions online allows other users to see changes that are going through the submission process. As with MeSH and EMTREE, the primary source of changes in to add new terms, however many changes are also related to creating or updating relations within GO (Klein).

To summarize, we see the following sources of change documented in the literature:
1. dealing with changing cultural and societal norms, specifically to address or correct bias;
2. political influence; and
3. new concepts and terminology arising from discoveries or change in perspective within a technical/scientific community.

# 3. New Sources of Change
The existing literature has focused primarily on sources of change stemming from practices associated with professional curators of KOSs. These include subject specialists, curators, or knowledge engineers. Two new actors, non-professional contributors (i.e. the crowd) and software, are now critical to the development of KOS whereas in prior generations they were somewhat ancillary. The involvement of these actors fundamentally changes how KOS are created and maintained.

## 3.1 Crowdsourcing
The emergence of Wikipedia and other crowdsourced based information systems has clearly impacted thinking behind the construction of KOS. (Vos 2006) described the differences between Wikipedia and Delicious categorizing systems and those of MeSH and the Dewey Decimal System. The bottom up style of the crowd sourced system is clearly evident and the overall network structure follows a stronger power-law like distribution.

(Suchecki et al. 2012) study the evolution of Wikipedia's category structure from its inception in 2004 till 2012. They show that the categorization system becomes stable over time especially with the introduction of top-level classification elements. Building on this work, (Bairi et al. 2015) also look at Wikipedia's evolution, they show that much of the evolution of Wikipedia's KO is about maintaining overall knowledge coherence rather than adding new knowledge. For example, there has been a 25% increase in the number of categories over the 2012 - 2014 period vs a 12% increase in the number of articles. Likewise, the number of disambiguation pages has increased by 13%. Both of these analyses point to change coming in the form not of just of additional concepts or categories, but change coming from ongoing maintenance that is much more active in bottom up derived KOS. This source of change is termed gardening.

While Wikipedia's knowledge organization does mirror that of more professionally curated systems (Suchecki et al. 2012), it is not captured with explicit semantics. That is, there is no official version defined using a formal representation language (e.g. SKOS or OWL). This is however beginning to change with the introduction of Wikidata (Vrandečić and Krötzsch 2014), this is beginning to change. Wikidata provides a structured data version of much of the information available within Wikipedia's infoboxes. The information is present in a standard instance and class hierarchy mirroring RDF(S). However, with more formal semantics applying them consistently becomes challenging with over 16,000 active contributors (https://www.wikidata.org/w/index.php?title=Wikidata:Statistics&oldid=320545760) . This is documented in the dicussion page around help for membership properties (https://www.wikidata.org/w/index.php?title=Help_talk:Basic_membership_properties&oldid=260792038) where Wikipedians discuss how to present help around how to interpret properties like 'instance of', 'sub class of' and 'part of'. Likewise, these more formal properties may be slower to be available across the totality of the KO. For example, property constraints are just currently being developed and are applied on only a small subset of Wikidata entity descriptions (Erxleben et al. 2014).

Wikidata also sees a speed of change in its knowledge that is at least an order of magnitude more than most traditional KOS. Since its inception it has had more than 350 million edits (https://www.wikidata.org/w/index.php?title=Wikidata:Statistics&oldid=320545760) this stems from both the number of contributors as well as major usage of automated agents. It is important to note that Wikidata like Wikipedia is systematically versioned.

Wikidata points at two new sources of change: 1) incremental and high speed modifications rather than sequential releases; 2) progressive formalization rather than consistent and well known formalization (one is not guaranteed that formal semantics is applied throughout and on all concepts).

While we have focused on Wikipedia and Wikidata as exemplars of the crowd construction of KOSs, this happens in other sites for example for books on the website Library Things (Heymann, P et al. 2010). Even if the construction of a KOS is not organized by the crowd, it is increasingly likely that the crowd will be part of its construction as will be discussed in the ASIS&T 2016 panel "Crowdsourcing Approaches for Knowledge Organization Systems: Crowd Collaboration or Crowd Work?".

## 3.2 Automated Knowledge Base Construction

As noted earlier, software and in particular automated extraction systems are now an important part of creating KOS. This is shown by the extensive use of bots in Wikidata (https://www.wikidata.org/wiki/Wikidata:Bots). Additionally, there are a wide variety of systems that automatically construct knowledge bases by both automated text extraction and data integration (Suchanek et al. 2014). It is worth noting that these system acquire both terminologies - the t-box - as well as statements conforming to the acquired knowledge organization - the a-box. There is a long history of automated knowledge classification, extraction and markup (Gangemi 2013).

However, this is an increasingly active field because of application of knowledge bases in large-scale search by the likes of Google and Microsoft under the heading of knowledge graphs (Dong, X. et al.

2014). According to (Nickel et al. 2016), these knowledge bases can contain billions of facts and thousands of types.

Automated knowledge base construction employs a variety of techniques, from open information extraction, to link prediction and data integration. Indeed, (Biega et al. 2013) detail the usage of over 30 different extraction algorithms by the Yago system. Moreover, these systems not only apply multiple algorithms but also use multiple sources that in turn are built of subsequent sources (Groth 2013). For example, the NELL system derives its knowledge organization by crawling millions of web pages constantly (Mitchell 2015).

These automated mechanisms for constructing KOSs introduce three important sources of change. The first source is that changes to algorithms can impact the resulting knowledge organization. Even if the source of data were to remain constant if a particular pipeline or extractor is changed, it can impact the results. This is not too dissimilar to the impact a change in knowledge engineer can have but at larger scale and in a more opaque and diffuse fashion.

The second source of change is the breadth of underlying data and it is (sometimes) unclear provenance that can be used to build a KOS. No longer is a KOS derived from multiply sourced scientific articles as in the case of MESH or GO but instead from the Web as whole or multiple independent sources. Furthermore, a KOS can be designed to constantly update itself as new sources become available. An important variant of this later source of change is the impact that underlying data can have when used to train algorithms, which are in turn used to build and maintain a KOS. Thus, as the source of training data changes so does the KOS.

To summarize, we see the following new sources of change:
1. gardening (i.e. ongoing maintenance);
2. incremental contributorship;
3. progressive formalization;
4. software and automation;
5. integration of large numbers of data sources; and
6. variance in algorithm training data.

## 4. Discussion

Our review of sources of change to KOSs and how people contribute change has revealed a few interesting patterns. Some of the older models like Dewey and LCSH were broad in scope and had wider distribution. Along with that, they approached change with a relatively conservative approach. This seems logical as the changes they were making affected many stakeholders and rolling back (restoring) changes would cause problems to downstream consumers. Also, as noted, many of the changes were regarding sensitive topics about people and culture.

As new KOs are developed, new significant sources of change to models are emerging: non-professional contributors and software. The positive aspects to harnessing these new sources is their volume and efficiency. There may also be a perception that these processes remove much of the human element from KOS design and the output is somehow neutral, unbiased, and accurate. The fact is that bias will continue to permeate through these processes. As an example, the uneven distribution of topics in Wikipedia could lead to bias appearing in a KOS that used it as a source. As far as accuracy goes, algorithms are dependent on quality training data and the best practices for developing training data is to have a quality selection workflow with heavy involvement from subject matter experts. This could ultimately result in another form of bias.

The KOS that result from crowdsourcing and software will not be absent from human supervision. However, the roles that humans play will be quite different from previous ones. As described earlier, the traditional KOS followed a production model similar to an editorial board with section specialists and subject matter experts. There will still be roles for specialists and experts in new models. The ideal ones will likely be hybrid models where software does much of the heavy lifting to detect and recommend new concepts. In these workflows, specialists will needed to verify recommendations and make sure that they are in scope of the model and the applications that are using it. Experts will also be needed to contribute to some linguistic aspects to the concepts. For example, experts will be needed to select preferred terms and synonyms.

If modern and future KOS development is going to involve a hybrid approach, it means that there are additional design considerations. When an expert is reviewing a suggestion for a change, they will want to know the source and provenance of the suggestion. This would likely start with categories of recommendations. Did the suggested concept come from an algorithm, a crowdsourced model, and a user suggestion? And from there, they will want to capture the specific source and record provenance information about the particular recommendation that will be saved with the metadata for that concept. Additional information will have to be carried along for suggestions that come from trained algorithms. Presumably, for an expert to receive a notification that a new concept needs to be reviewed, that recommendation would have had to pass a certain 'likelihood' threshold. The editor would want the ability to review the confidence score for that concept and put it into context other similarly suggested terms.

## 5. Conclusion

We have cataloged nine sources of change that impact the construction and design of knowledge organization systems: changing norms, political influence, new developments, maintenance; incremental contributorship; progressive formalization; software and automation; integration of large numbers of data sources; and variance in algorithm training data.

Six of these changes are largely the result of new mechanisms for KOS construction, in particular, the introduction of crowdsourcing and automation. We hope that this list helps those responsible for designing, building and maintaining KOSs reflect on appropriate policies, guidelines and development practices to deal with change.